\documentclass[aps,preprint,showpacs,floatfix]{revtex4}
\usepackage{graphicx,bm,amsmath,amssymb,dcolumn}
\begin{document}
\newcommand{\half}{\frac{1}{2}}
\newcommand{\ith}{^{(i)}}
\newcommand{\im}{^{(i-1)}}
\newcommand{\br}{{\mathbf r}}
\newcommand{\ba}{{\mathbf a}}
\newcommand{\bb}{{\mathbf b}}
\newcommand{\Ham}{\beta {\mathcal H}}
\newcommand{\cE}{{\mathcal E}}
\newcommand{\tsigma}{\tilde{\sigma}}
\newcommand{\tv}{\tilde{v}}
\newcommand{\ahat}{{\widehat{a}}}
\newcommand{\xhat}{{\widehat{x}}}
\newcommand{\yhat}{{\widehat{y}}}
\newcommand{\vecr}{\vec{r}}
\newcommand{\veca}{\vec{a}}
\newcommand{\vecb}{\vec{b}}
\newcommand{\gae}
{\,\hbox{\lower0.5ex\hbox{$\sim$}\llap{\raise0.5ex\hbox{$>$}}}\,}
\newcommand{\lae}
{\,\hbox{\lower0.5ex\hbox{$\sim$}\llap{\raise0.5ex\hbox{$<$}}}\,}
\newcommand{\uptr}{\hbox{\raise0.0ex\hbox{$\bigtriangleup$}}}
\newcommand{\dntr}{\hbox{\raise0.5ex\hbox{$\bigtriangledown$}}}
\title {
Phase transitions in self-dual generalizations of the Baxter-Wu model}
\author{
Youjin Deng$^1$, Wenan Guo$^2$, Jouke R. Heringa$^3$,
Henk W. J. Bl\"ote$^4$ and Bernard Nienhuis$^5$ }

\affiliation{$^{1}$ Hefei National Laboratory for Physical
Sciences at Microscale,
Department of Modern Physics, University of Science and
Technology of China, Hefei 230027, China }
\affiliation{$^{2}$Physics Department, Beijing Normal University,
Beijing 100875, P. R. China }
\affiliation{$^{3}$ Fundamental Aspects of Energy and Materials,
Faculty of Applied Sciences, Delft University of
Technology, Mekelweg 15, 2629 JB Delft, The Netherlands}
\affiliation{$^{4}$ Instituut Lorentz, Leiden University,
  P.O. Box 9506, 2300 RA Leiden, The Netherlands}
\affiliation{$^5$ Instituut voor Theoretische Fysica, Universiteit van
Amsterdam, Valckenierstraat 65, 1018 XE Amsterdam, The Netherlands }
\date{\today}
\begin{abstract}
We study two types of generalized Baxter-Wu models, by means of
transfer-matrix and Monte Carlo techniques. The first generalization
allows for different couplings in the up- and down triangles, and 
the second generalization is to a $q$-state spin model with
three-spin interactions. Both generalizations lead to self-dual
models, so that the probable locations of the phase transitions
follow. Our numerical analysis confirms that phase transitions
occur at the self-dual points. For both generalizations of the Baxter-Wu
model, the phase transitions appear to be discontinuous.
\end{abstract}
\maketitle
\section{Introduction}
\label{intro}
In general, systems in the universality class of the two-dimensional
4-state Potts model display critical singularities that are modified
by logarithmic correction factors. A satisfactory explanation of this
fact is provided by the renormalization scenario due to Nienhuis et
al.~\cite{NBRS}. It explains the logarithmic factors \cite{NS} as arising 
from the second temperature field, which is marginally irrelevant.
It also shows that the 4-state Potts behavior without
logarithmic factors can only occur at special points in the parameter
space, where the two leading temperature fields simultaneously vanish. 
The exactly solved Baxter-Wu model \cite{BW} precisely fits such
a location in parameter space: it belongs to the 4-state Potts class
and its leading critical singularities do not have logarithmic factors.
Its reduced Hamiltonian reads 
\begin{equation}
\Ham = -K^{\rm I}  \sum_{\uptr \dntr}
 s_i s_j s_k 
\label{BW}
\end{equation}
where $\beta=1/(k_BT)$ is the inverse temperature, and
the sum is over the up- and down triangles of the triangular
lattice, and the site labels $i$, $j$ and $k$ refer to the three spins
at the vertices of each triangle. Each spin assumes the Ising values
$\pm 1$; this is emphasized by the superscript I of the coupling
$K^{\rm I}$.  At low temperatures, the model is in one of four long-range
ordered phases, where most triangles have an even number of $-$ spins.
While the common type of interaction between spins
in magnetic materials is of the two-spin type, three-particle interactions
such as in the Baxter-Wu model have been used to describe the
shape of face-centered cubic crystal surfaces \cite{BaN}.

\begin{figure}
\includegraphics[scale=1.2]{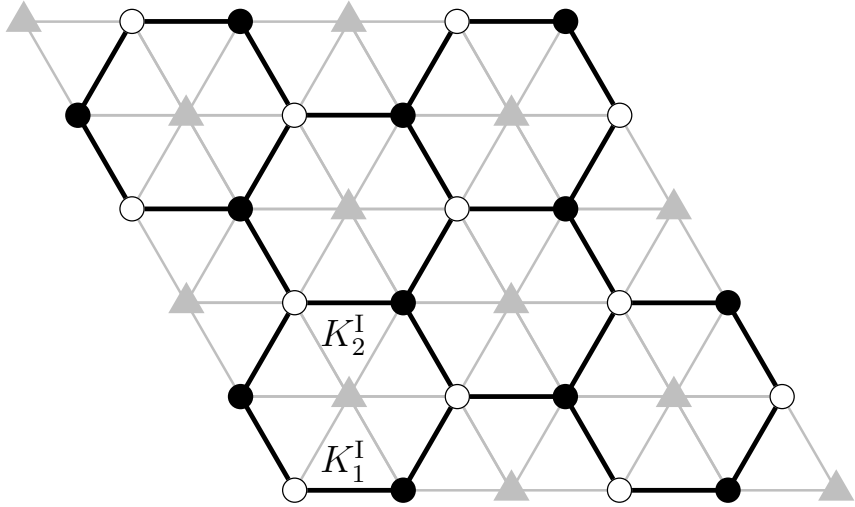}
\caption{
A $6 \times 6$ triangular lattice. The lattice is divided into a 
honeycomb (open and filled circles) and a triangular sublattice (pentagons), 
which are dual to each other. The honeycomb lattice is bipartite.}
\label{triangle}
\end{figure}

This work investigates two different generalizations of the Baxter-Wu
model. First we consider the case that the couplings in the up- and
down triangles are different (see Fig.~\ref{triangle}), i.e.,
\begin{equation}
\Ham = -K^{\rm I}_1 \sum_{\uptr} s_i s_j s_k  
       -K^{\rm I}_2 \sum_{\dntr} s_i s_j s_k
\label{BW2c}
\end{equation}
where the sums are over the up- and down triangles of the triangular
lattice respectively. The introduction of another temperature-like
parameter makes it likely that this model will have a critical line 
parametrized by the ratio of $K^{\rm I}_1$ and $K^{\rm I}_2$. The fourfold
degeneracy of the ground state persists for $K^{\rm I}_1 \ne K^{\rm I}_2$,
so that it may seem plausible that the model still belongs to
the 4-state Potts universality class. We shall attempt to provide a
more definite judgment by means of a numerical investigation.

For the second generalization it is useful to write Eq.~(\ref{BW})
in terms of two-state Potts variables $\sigma_i\equiv (s_i+3)/2=1$ or 2:
\begin{equation}
\Ham = -K  \sum_{\uptr \dntr}
\delta_2 (\sigma_i+\sigma_j+\sigma_k)
\label{BW2}
\end{equation}
where $\delta_2 (x)=0$ if $x$ is odd and 1 if $x$ is even, and 
$K=2K^{\rm I}$. The sum 
is over all up- and down triangles. Eqs.~(\ref{BW}) and (\ref{BW2})
differ by an additive constant that is irrelevant for the present
purposes.
It is now straightforward to generalize the model in terms of $q$-state
variables with values $\sigma_i=1,~2,~\cdots ,~q$:
\begin{equation}
\Ham = -K  \sum_{\uptr \dntr}
\delta_q (\sigma_i+\sigma_j+\sigma_k)
\label{BWq}
\end{equation}
where $\delta_q (x)=1$ if $(x \mod q)=0$ and $\delta_q (x)=0$ otherwise.
This model can also be considered as a generalization of the $q$-state Potts
model  \cite{RBP} to 3-spin interactions, because the pair couplings of
the original Potts model on a bipartite lattice can be written as
$ -K\delta_q (\sigma_i+\sigma_j)$. But the model (\ref{BWq}) does not
obey the $q$-fold permutation symmetry ${\mathcal S}_q$ of the Potts
model for general $q$.  Its symmetry group is 
${\mathcal Z}_q\otimes{\mathcal Z}_q\otimes{\mathcal Z}_2\otimes{\mathcal S}_3$
where the $q$-state clock symmetries ${\mathcal Z}_q$ are generated by the
operation $\sigma_i \to \sigma_i +1 \mod q$, independently for two of the
three sublattices, ${\mathcal Z}_2$ is generated by the operation 
$\sigma_i \to q+1-\sigma_i$ on
all sites, and ${\mathcal S}_3$ is the symmetric group of the permutations
of the three sublattices. The latter symmetry results from the spatial
symmetries of the lattice, namely reflection and translation or rotation,
which can permute the three sublattices, while leaving $\Ham$ invariant.

It is obvious that the degeneracy of the ground state increases
as $q^2$ with the number $q$ of spin states, so that one may expect that
the model will display a discontinuous ordering transition for $q>2$.
However, the special nature of the critical Baxter-Wu model, i.e.
the model of Eq.~(\ref{BWq}) for $q=2$, namely the vanishing
of the marginal temperature field, opens the possibility of another
scenario. After a mapping on the Coulomb gas \cite{BN}, the marginal
temperature field translates into the fugacity of the $e=4$ electric
charges. Thus the $q=2$ transition maps precisely on the point of
the Gaussian fixed line where the electric charges are absent, and
there seems to be a real possibility that this is also the case for
other values of $q$. Since one expects that the Coulomb gas coupling
increases with $q$, the electric charges, which are marginal at $q=2$,
must be relevant for $q>2$, and would drive the ordering transition 
first order. But, if these charges remain absent, the transition still
takes place on the Gaussian line, and must be critical.

For this reason it is interesting to investigate the character of the
ordering transition for $q>2$. There are existing results due to
Alcaraz et al.~\cite{AJ,ACO} who investigated a different generalization
of the Baxter-Wu model, namely, to a $p$-state clock model. For the
case $p=3$, their model is equivalent with our $q=3$ model. They
concluded that the transition is first-order for $p=3$, on the basis
of approximate renormalization calculations, and Monte Carlo
calculations starting in the ordered and the disordered states,
displaying changes of phase.

The property of self-duality plays an important role in the present
work, because knowledge of the critical point greatly facilitates
the numerical analyses.  Its derivation is the subject of 
Sec.~\ref{dual} where we formulate
a relatively simple proof of self-duality for a class of
models that includes both generalizations of the Baxter-Wu model
mentioned above.
In Sec.~\ref{2coup} we present our numerical analysis of the
$q=2$ model with two different couplings, and in Sec.~\ref{qgt2}
we report our findings for the $q=3$ and 4 models with uniform
couplings. The conclusions
of our analyses are listed in Section \ref{sec_con}.

\section{Duality of $q$-state models with multispin interactions}
\label{dual}
Self-duality is a useful tool to locate phase transitions.
If a single phase
transition occurs as a function of temperature, then the transition
must occur at the point where the temperature variable $K$
and the dual temperature variable $\tilde{K}$ coincide.
In the case of self-dual models with two variables $K_1$ and
$K_2$, the transitions tend to occur on the  self-dual line
in the $K_1,K_2$ plane, i.e., in a point that maps 
onto itself under duality.

Duality was first found for the square-lattice Ising model by
Kramers and Wannier \cite{KraWa}, who correctly predicted the
critical point at
\begin{equation}
K^{\rm I}_{\rm c} = \half \ln (1+\sqrt{2})
\label{sdc}
\end{equation}
and since then many more
derivations have been reported. Gruber et al.~\cite{GHM} have
formulated a very general proof that includes all systems
studied in the present work. For the convenience of the reader
we shall provide a simple proof that is less general than that
of Gruber et al.~\cite{GHM}, but still more general than 
actually required for the models under the present investigation.

Simpler, and less general versions of the proof given by Gruber et al.
appear elsewhere in the literature. Examples are the two-dimensional Ising
model with pair interactions in one direction and multispin interactions
in the perpendicular direction (see Refs.~\cite{Turban}, and
\cite{ZY} for a generalization to $q>2$ Potts models with
similar interactions).

The present derivation of self-duality applies to a system of $q$-state
variables located on a simple hypercubic lattice. The variables are
denoted $\sigma_{\br}$ and take the values $1,\cdots ,q$.
Their interactions are described by a Hamiltonian of
the general form
\begin{equation}
- \Ham  = K_1\sum_{\br}
 \delta_q \left( \sum_{i=1}^n \sigma_{\br+\ba_i} \right)
 + K_2\sum_{\br}
 \delta_q \left( \sum_{j=1}^m \sigma_{\br+\bb_j} \right)
\label{eq_Ham}
\end{equation}
where $\br$ is a lattice vector and $\ba_i$ and $\bb_j$
are vectors pointing from position $\br$ to the sites of the 
variables participating in the interaction assigned to site $\br$.
There are two multiparticle interactions per site, one with $n$
participating sites and another with $m$ sites.
The class includes the square-lattice Potts model with nearest-neighbor
interactions, after a suitable renaming $\sigma \to q-\sigma$ of the
$q$ states on one of the two sublattices. It also includes the
Baxter-Wu model for $n=m=3$, $a_1=-b_1=(0,0)$,  $a_2=-b_2=(1,0)$, 
$a_3=b_3=(0,1)$, $q=2$, and $K_1=K_2$. 
The partition function for our class of models takes the form:
\begin{equation}
Z(v_1,v_2)=\sum_{\{\sigma_{\br}\}}\prod_{\br}
\left[1+v_1 \delta_q \left(\sum_{i=1}^n \sigma_{\br+\ba_i}\right) \right]
\left[1+v_2 \delta_q \left(\sum_{j=1}^m \sigma_{\br+\bb_j}\right)\right],
\label{eq-part1}
\end{equation}
where $v_1=\exp (K_1)-1$ and $v_2=\exp (K_2)-1$.
Each $\delta_q$-function in Eq.~(\ref{eq-part1}) can be substituted
by its Fourier representation
\begin{equation}
\delta_q (z) =\frac{1}{q} \sum_{t=1}^q e^{\pm 2\pi i z t/q}    \, ,
\label{Four_tran}
\end{equation}
and each ``1" in Eq.~(\ref{eq-part1}) can be replaced using the identity    
\begin{equation}
1 =  \sum_{t=1}^q \delta_q (t) e^{\pm 2\pi i z t/q}  \; .    
\label{eq-identity}
\end{equation}
The effect of these substitutions is that two new variables $t_{\br}$ 
and $t'_{\br}$ are introduced on each site $\br$, for the $n$- and
$m$-particle interactions, respectively.  This leads to
\begin{eqnarray}
Z (v_1,v_2) & = &  \sum_{\{\sigma\}} \prod_{\br} \sum_{\{t_{\br},t'_{\br}\}}
[\delta_q ( t_{\br})+ \frac{v_1}{q} \, ]
[\delta_q (t'_{\br})+ \frac{v_2}{q} \, ]  \nonumber \\
&&
 \exp \left[ \frac{2 \pi i }{q }
\left( t_{\br} \sum_{i=1}^n \sigma_{\br +\ba_i} -
t'_{\br} \sum_{j=1}^m \sigma_{\br +\bb_j}\right) \right]  \; .
\end{eqnarray}

After reordering the summations and the products and collecting terms
with the same $\sigma$, we obtain
\begin{eqnarray}
Z (v_1,v_2) & = & (v_1 v_2/q)^{N} \sum_{\{t,t' \}} \prod_{\br} 
[1+(q/v_1) \delta_q (t_{\br})] [1+(q/v_2)\delta_q (t'_{\br})]  \nonumber \\
&& \sum_{\{ \sigma \}}
(1/q) \exp \left[ (2 \pi i \sigma_{\br}/q )
\left( \sum_{i=1}^n t_{\br -\ba_i} -
\sum_{j=1}^m t'_{\br -\bb_j}\right) \right]  \; ,
\label{Four_Z_Pott}
\end{eqnarray}
where $N$ is the total number of sites in the lattice.
A nice property of Eq.~(\ref{Four_Z_Pott}) is that the degrees freedom  
$\sigma_{\br}$ on different sites $\br$ are completely independent, and 
thus the summation over the $\sigma_{\br}$ becomes very easy. Using again 
Fourier-transformation~(\ref{Four_tran}), one has
\begin{eqnarray}
Z(v_1,v_2)&=& (v_1 v_2/q)^N \sum_{\{t_{\br},t'_{\br}\}}\prod_{\br}
[1+(q/v_1) \delta_q (t_{\br})] [1+(q/v_2)\delta_q (t'_{\br})]  \nonumber \\
&& \delta_q \left( \sum_{i=1}^n t_{\br-\ba_i} - \sum_{j=1}^{m}
t'_{\br-\bb_j}\right) \; .
\label{eq-part2}
\end{eqnarray}
In short, the original $q$-valued variable $\sigma_{\br}$ has been 
integrated out.
The price paid is the introduction on each site of two new $q$-valued
variables $t_{\br}, t'_{\br}$ 
with an additional $\delta$-function constraint. 

Next, one introduces a new $q$-state variable $\tsigma_{\br}$ on each site,
and let $t$:
\begin{equation}
t_{\br}=\sum_{j=1}^m \tsigma_{\br-\bb_j} \bmod q,
\end{equation}
which will be feasible for appropriate boundary conditions.
The $\delta$ function connecting  $t_{\br}$ and $t'_{\br}$
in Eq.~(\ref{eq-part2}) is satisfied if
\begin{equation}
t'_{\br}=\sum_{i=1}^n \tsigma_{\br-\ba_i} \bmod q \; .
\end{equation}
As the number of new variables $\tsigma_{\br}$ is equal to the number of
old variables $t$ and $t'$ reduced by the number of constraints on $t$ and
$t'$ imposed by the rightmost $\delta$ function in Eq.~(\ref{eq-part2}),
we expect that the $\tsigma_{\br}$ are determined up to a trivial shift.
After an inversion of the lattice,
Eq.~(\ref{eq-part2}) takes the form 
\begin{equation}
Z(v_1,v_2) = \left(\frac{v_1 v_2}{q}\right)^N
\sum_{\{\tsigma_{\br}\}}\prod_{\br}    
\left[1+ \frac{q}{v_2} \,
\delta_q \left(\sum_{i=1}^n \tsigma_{\br+\ba_i}\right)\right]
\left[1+ \frac{q}{v_1} \,
\delta_q \left(\sum_{j=1}^m \tsigma_{\br+\bb_j}\right)\right],
\label{eq-part3}
\end{equation}
Comparison with Eq.~(\ref{eq-part1}) shows that $Z(v_1,v_2)$ satisfies the
self-duality relation
\begin{equation}
Z(v_1,v_2)=\left(\frac{v_1 v_2}{q}\right)^N 
Z(\tv_1,\tv_2) \hspace{2cm} \mbox{with} \hspace{0.5cm}
v_1 \tv_2 =q \, , ~~  v_2 \tv_1 =q \; .
\label{eq_sd}
\end{equation}
The dual set of coupling constants $(\tilde{K}_1,\tilde{K}_2)$ obey
\begin{equation}
\tv_1 =e^{\tilde{K}_1}-1  \hspace{1cm} \mbox{and} \hspace{1cm}
\tv_2 =e^{\tilde{K}_2}-1  \; .
\label{eq_sd2}
\end{equation}
Each point on the line
\begin{equation}
v_1 v_2 =q
\label{Psdline}
\end{equation}
is mapped onto itself, and we find, for the case $v_1=v_2$ 
the symmetric self-dual point as $v=\sqrt{q}$ or
\begin{equation}
K=\ln(1+\sqrt{q}) \, .
\end{equation}
In this self-dual point the average number of satisfied multiparticle
interactions (``satisfied'' means that the sum modulo $q$ of the spins 
coupled by the interaction vanishes) per site, if unique, is found from
the derivative of $\ln Z$ with respect to the coupling constants at
the self-dual point.
In the case of a first-order transition on the self-dual line, this
yields the mean of the values in the disordered phase and
in the ordered phase.

For $q=2$ models defined in terms of Ising spins $s_i= \pm 1$, one
has to take into account the factor 2 between the ``Potts'' and
``Ising'' couplings, as appearing under Eq.~(\ref{BW2})--i.e.,
$v=\exp (2K^{\rm I})-1$. 
In the Ising case,
the equation for the self-dual line Eq.~(\ref{Psdline}) may be
written as
\begin{equation}
\sinh (2K^{\rm I}_1) \sinh (2K^{\rm I}_2) = 1
\label{Isdline}
\end{equation}

In many cases, the self-dual line, or a part of it, is the locus of
a phase transition. The existence, uniqueness, and character of a 
phase transition, however, are not determined by self-duality.
For that purpose, additional calculations are required.
For several Ising models with multispin interactions and a field
($m=1$), including three-dimensional models, Bl\"ote et al.~\cite{sedua} 
found discontinuous transitions on a part of the self-dual line,
with a gas-liquid like critical point at the end of the first-order
range.
For a two-dimensional system with pair interactions in one direction
and multiparticle interactions between $p$ particles in the perpendicular
direction, Zhang and Yang \cite{ZY} concluded, from Monte Carlo
calculations, that a phase transition occurs at the self-dual point,
and that it is first-order for all $q>2$ if $p>2$.
Also in the case of the $n$-state clock model with three-particle
interactions on the triangular lattice, Alcaraz et al.~found 
from Monte Carlo calculations \cite{AJ} that phase transitions occur
at the self-dual points for $n=2$ and $n=3$.

\section{Numerical Methods}
We investigate the generalized Baxter-Wu model~(\ref{eq_Ham}) on the
triangular lattice, both by transfer-matrix method and by Monte Carlo
simulations.
\subsection{Transfer-matrix }
\label{2coupTM}
The transfer-matrix techniques used in this work are adequately described
in the literature, although the information is divided over different
papers. The essential parts are explained in Refs.~\cite{MPN},
\cite{BN1982} and \cite{QWB}. Here we only add a few
general and specific remarks for the convenience of the reader.
From a few of the leading eigenvalues of the transfer matrix, one can
calculate the free energies, the magnetic and energy-like correlation
lengths of $L \times \infty$ systems. For the case $q=2$ we could 
perform such calculations up to finite sizes $L=27$.
The geometry is that of the triangular lattice wrapped on a cylinder,
with one set of edges perpendicular to the axis of the cylinder.
The finite size $L$ is specified such that the circumference of the
cylinder is spanned by $L$ lattice edges.

Here we use the true triangular lattice, instead of the representation
as a square lattice with one set of diagonal bonds, as used in
Sec.~\ref{dual}. Since, after adding one layer of spins, the lattice
is shifted by a half lattice unit along the finite direction, we chose a
transfer matrix that adds two layers of spins and applies an additional
reverse shift operation, in order to ensure that the transfer matrix
commutes with the lattice reflection as specified below. Such commutation
relations allow one to find a common set of eigenstates of the transfer
matrix and a symmetry operator.

The transfer matrix acts on a vector space with vector indices
representing the state of a row of $L$ Ising spin variables. 
For $q=2$, the vector indices can thus be written as binary
numbers $b_L b_{L-1}\cdots b_2 b_1$ with $b_k\equiv (s_k+1)/2$.
For $q=3$ one uses ternary numbers, etc., but here we shall use the
language for binary numbers.  The transfer matrix calculations
focus on three eigenvalues, namely the largest one $\lambda_0$, the
"magnetic" one $\lambda_m$, and the "thermal" eigenvalue $\lambda_t$. 
These eigenvalues are defined in the usual way,
by means of the group of symmetry operations that leave the Hamiltonian
invariant, but permute the ordered phases.  The thermal eigenvalue, like
the largest eigenvalue corresponds to an eigenvector fully invariant
under these symmetry operations.  The magnetic eigenvalue is the largest
one with an eigenvector that changes under these symmetry operations.
In this model the relevant symmetry group is generated by the allowed
permutations of the $q$ states, and by lattice symmetries that permute
the three sublattices.  As the transfer matrix breaks some of the
latter symmetries, we replace the full symmetry group by the subgroup
that is not violated by the transfer matrix.

The analyses based on  $\lambda_t$ and  $\lambda_m$ are similar.
We proceed as follows for the case of $\lambda_m$.
The magnetic correlation function $g_{m}(r)$ as a function of the 
distance $r$ in the length direction of the cylinder is defined as
$g_{m}(r) = \langle s_{0}s_{r}\rangle$.
For sufficiently large $r$, $g_{m}(r)$ decays exponentially on a
length scale $\xi_m$ that depends on $L$ and the couplings, i.e.,
\begin{equation}
g_{m}(r) \propto {\rm e}^{-r/\xi_m(K_1,K_2,L)}
\end{equation}
and is determined by the eigenvalues $\lambda_0$ and $\lambda_m$ of the
transfer matrix:
\begin{equation}
\xi_m^{-1}(K_1,K_2,L) =
\frac{1}{\sqrt{3}} \ln(\lambda_0/\lambda_m) \, . 
\label{calxi}
\end{equation}
The geometric factor $\sqrt{3}$ allows for the thickness of two layers
added by the transfer matrix, expressed in the same unit as the finite
size $L$.  With the help of Cardy's conformal mapping \cite{Cardy-xi} of
the infinite plane on a cylinder with a circumference $L$, one can now,
for a system at criticality, relate the magnetic scaling dimension $X_h$,
which describes the algebraic decay of the correlation function in the
infinite system, to $\xi_m$. Defining the scaled gap $X_{h}(K_1,K_2,L)$ by
\begin{equation}
X_{h}(K_1,K_2,L) \equiv \frac{L}{2 \pi \xi_m(K_1,K_2,L)} \, ,
\label{calsg}
\end{equation}
and using finite-size scaling \cite{FSS}, one finds that, at criticality,
\begin{equation}
X_{h}(K_1,K_2,L)= X_{h} +b_1 L^{y_1} + b_2 L^{y_2} + \cdots
\label{Xscal}
\end{equation}
where the correction terms $b_i L^{y_i}$ arise from irrelevant fields,
whose presence means that conformal invariance applies only in the
limit of large length scales. Since the irrelevant exponents satisfy
$y_i<0$, $X_{h}(K_1,K_2,L)$ converges to $X_{h}$ with increasing $L$,
and numerical estimates of $X_{h}$ can be obtained from the finite-size
data that can be calculated for a range of system sizes. 

For a system that is not critical due to the presence of some relevant
scaling field, a term with a positive power of $L$ appears in
Eq.~(\ref{Xscal}), which will lead to crossover to different behavior,
for instance described by a zero-temperature or an infinite-temperature
fixed point. A finite-size analysis of the quantity $X_{h}(K_1,K_2,L)$
may thus show whether or not the system is critical, and if so,
provide information on the universality class of the model.

The analysis of the temperature dimension $X_t$ from the energy-like
correlation length $\xi_t$ similarly uses the  eigenvalue $\lambda_t$.
The calculation of this eigenvalue, with the same symmetry as
$\lambda_0$, is described in Ref.~\cite{BN1982}.

\subsection{Monte Carlo algorithm}
\label{2coupMC}
Simulation of the generalized Baxter-Wu model on the triangular lattice
can simply employ the standard Metropolis method which involves 
single-spin updates only. However, a more efficient 
algorithm--a Swendsen-Wang-type cluster Monte Carlo 
method--can be formulated, which was already described for
the Baxter-Wu model in Ref.~\cite{NE}.

To construct such a cluster method, one first divides the triangular
lattice $\mathcal T$ into three sublattices ${\mathcal L}_{T_1}$,
${\mathcal L}_{T_2}$, and ${\mathcal L}_{T_3}$ which are
triangular. The union of any two sublattices
form a honeycomb lattice ${\mathcal L}_H$ which is dual to the remaining
triangular lattice (see Fig.~\ref{triangle}). 
The partition sum of a generalized Baxter-Wu model can then be written
\begin{equation}
Z(v_1,v_2) = \sum_{\{ \sigma_k \}} \sum_{\{ \sigma_i,\sigma_j \}} 
\prod_{\langle ij \rangle}  e^{ K_1 
\delta_q (\sigma_i +\sigma_j+\sigma_k)}  
e^{K_2 \delta_q (\sigma_i +\sigma_j +\sigma_{k'})}
\label{Par_SW}
\end{equation}
where the product is over every edge of the honeycomb sublattice 
${\mathcal L}_H$, and $k$ and $k'$ are the two neighboring sites on the
remaining triangular sublattice, on either side of edge $\langle ij \rangle$.
The statistical weight associated with each edge $\langle ij \rangle$ 
is then
\begin{eqnarray}
& &e^{K_1   \delta_q (\sigma_i +\sigma_j +\sigma_k)}
   e^{K_2   \delta_q (\sigma_i +\sigma_j +\sigma_{k'})}=    \nonumber \\
& &  [1+v_1 \delta_q (\sigma_i +\sigma_j +\sigma_k) ]  
     [1+v_2 \delta_q (\sigma_i +\sigma_j +\sigma_{k'})]=    \nonumber \\
& &  \sum_{b_{ij}^{(1)}=0,1}
       [v_1 \delta_q (\sigma_i +\sigma_j +\sigma_k)]^{b_{ij}^{(1)}}
      \sum_{b_{ij}^{(2)}=0,1}
       [v_2 \delta_q (\sigma_i +\sigma_j +\sigma_{k'})]^{b_{ij}^{(2)}}
\label{edge_weight}
\end{eqnarray}
where
$v_1 =\exp (K_1)-1$ and $v_2 =\exp (K_2)-1$,
and the convention $0^0=1$ has been used. Thus, by introducing two bond
variables $b_{ij}^{(1)}, b_{ij}^{(2)}$ for every edge of ${\mathcal L}_H$,
and replacing the corresponding edge weights in Eq.~(\ref{Par_SW})
according to Eq.~(\ref{edge_weight}), one obtains a joint spin-bond
model.

The Swendsen-Wang-type cluster method can be adapted to simulate the
joint spin-bond model. Two basic steps are involved: the bond- and the
spin updates. Given a spin configuration, Eq.~(\ref{edge_weight}) tells
that the bond updates can be performed as in a {\it uncorrelated} 
bond percolation: the bond-occupation probability is $p=v_1/(1+v_1)$ for
$b_{ij}^{(1)}$ on each edge with a satisfied up triangle, and 
$p=v_2/(1+v_2)$ for  $b_{ij}^{(2)}$ on each edge with a satisfied down
triangle, and $p=0$ otherwise.  Given a bond configuration, 
Eq.~(\ref{edge_weight}) tells that spin configurations satisfying the
$\delta$ functions have equal probability.  Making use of the fact that
the honeycomb lattice is bipartite, one can formulate the following
algorithm.\\
Cluster algorithm, {\bf version 1}:
\begin{enumerate}
\item {\it Sublattice division}. Randomly  with equal probability label
the three sublattices as 1, 2 and 3. Then merge two sublattices into
a honeycomb lattice 
${\mathcal L}_H \equiv {\mathcal L}_{T_2} \cup {\mathcal L}_{T_3}$.
\item {\it Bond update}. On each edge $\langle ij \rangle$ of the honeycomb 
sublattice ${\mathcal L}_H$, place an occupied bond with probability
$p= 1-e^{-K_1-K_2}$ if both the up- and the down-triangles are satisfied,
$p=1-e^{-K_1}$ if only the up triangle is satisfied, $p=1-e^{-K_2}$
if only the down triangle is satisfied, and $p=0$ otherwise.
\item {\it Cluster construction}. A cluster is defined as a group of
sites connected through occupied bonds, irrespective of colors.
Decompose the lattice ${\mathcal L}_H$ into clusters 
(including single-site clusters).
\item {\it Spin update}. All the spins on the triangular sublattice
${\mathcal L}_{T_1}$ are left unchanged. Randomly with uniform
probability choose a value
$\tau =0,1,2,\cdots,q-1$. Independently for each cluster, update the 
spins on sublattice ${\mathcal L}_{T_2}$
according to $\sigma \rightarrow (\sigma + \tau) \mbox{ mod } q$, and
the spins on ${\mathcal L}_{T_3}$ according to 
$\sigma \rightarrow (\sigma + q- \tau) \mbox{ mod } q$.
\end{enumerate}
This completes one Swendsen-Wang-type cluster step, and a new spin
configuration is obtained. Other choices are possible to choose $\tau$
in step 4, for instance $\tau =0$ with probability 1/2 and the other
values of $\tau$ with probability $1/(2q-2)$. The choice $\tau =0$
with probability 0 and the other values with probability $1/q$ is only
applicable for $q>2$.

For the special case $q=2$ the cluster algorithm can be made more
efficient. Conditional on the frozen spin configuration on sublattice
${\mathcal L}_{T_1}$, the honeycomb sublattice of the $q=2$ 
generalized Baxter-Wu model reduces to an Ising model with 
position-dependent couplings
on the honeycomb lattice  ${\mathcal L}_H$:
\begin{equation}
\Ham |_{\{ \sigma_k \}, k \in {\mathcal L}_{T_1} } = -
\sum_{\langle ij \rangle}s_i s_j (K_1^{\rm I}s_k+K_2^{\rm I}s_{k'})
\equiv -\sum_{\langle ij \rangle} K_{ij}s_i s_j \hspace{1cm}(s=\pm 1) 
\label{embedding_Ising}
\end{equation}
where the meaning of $k$ and $k'$ is the same as in Eq.~(\ref{Par_SW}).
The effective coupling $K_{ij}$ is defined by the right-hand side of
this equation, and can be ferromagnetic or
antiferromagnetic, depending on the spin variables $s_k$ and $s_{k'}$.
On the basis of Eq.~(\ref{embedding_Ising}), the ``{\it bond-update}''
step can be reformulated as follows.\\
Cluster algorithm, {\bf version 2}:
\begin{enumerate}
\setcounter{enumi}{1}
\item {\it Bond-update}. On each edge $\langle ij \rangle$ of
${\mathcal L}_H$, place an occupied bond with probability
$p=\max [0, 1-\exp (-2 s_i s_j K_{ij})]$.
\end{enumerate}
The other steps are equal to those of version 1.
An occupied bond can be either ``ferromagnetic'' or ``antiferromagnetic''
(between spins of opposite signs).  A cluster in version 1 may be
further decomposed into several clusters in version 2.

We found that version 2 performs much better than the
Metropolis algorithm, in the sense that a simulation using the cluster
method yields statistically more accurate results in a given time.
For the $q=2$ case with $K_1=K_2$, we found the dynamic exponent $z$
as about $1.1$, which is close to the Li-Sokal bound \cite{Li_89}
$z \geq 2 y_t -1 =1$. For the self-dual points with $K_1 \neq K_2$,
as well as those with $q >2$, a further increase of the slowing down
was observed.

We mention that a single-cluster version of the algorithm can also
be formulated. However, we found that it does not further improve the
efficiency. In fact, for the $q=2$ case with $K_1=K_2$, the dynamic   
exponent appears to exceed that of the full cluster-decomposition method.

\section{Results for $q=2$ and $K_1 \ne K_2$ }
\label{2coup}
For the present case $q=2$  we use the Ising notation for the
condition of self-duality as expressed by Eq.~(\ref{Isdline}).
Our numerical analysis divides into two parts. The transfer-matrix
results are described in subsection \ref{2coupTMres}.
The Monte Carlo investigation is reported in subsection \ref{mcrq2}.

\subsection{Transfer-matrix results }
\label{2coupTMres}
We calculated the scaled gaps at the self-dual points with
$K^{\rm I}_1=K^{\rm I}_2$, and $K^{\rm I}_1=0.5$, 0.6, $\cdots$, 1.2,
for system sizes up to $L=27$. The system sizes were restricted to
multiples of 3, because otherwise three of the four ground states
do not fit in a lattice with period $L$.
For the pure Baxter-Wu model at criticality, with
$K^{\rm I}_1=K^{\rm I}_2=[\ln(1+\sqrt{2})]/2$, we find that the
finite-size data for the scaled gaps rapidly approach the exact
values $X_h=1/8$ and $X_t=1/2$. Three-point fits according to
\begin{equation}
X_{h}(L) \simeq X_h +a L^{p} 
\label{3pfit}
\end{equation}
followed by iterated fits as described in Ref.~\cite{BN1982}
reproduce the exact values up to about $10^{-7}$.
For $K^{\rm I}_1 \ne K^{\rm I}_2$, the finite-size dependence of the
scaled gaps becomes stronger while the signs of convergence disappear,
at least for a certain range of $K^{\rm I}_1/K^{\rm I}_2$.
This is illustrated by the finite-size data in Table \ref{tab_1}.
\begin{figure}
\includegraphics[scale=0.9]{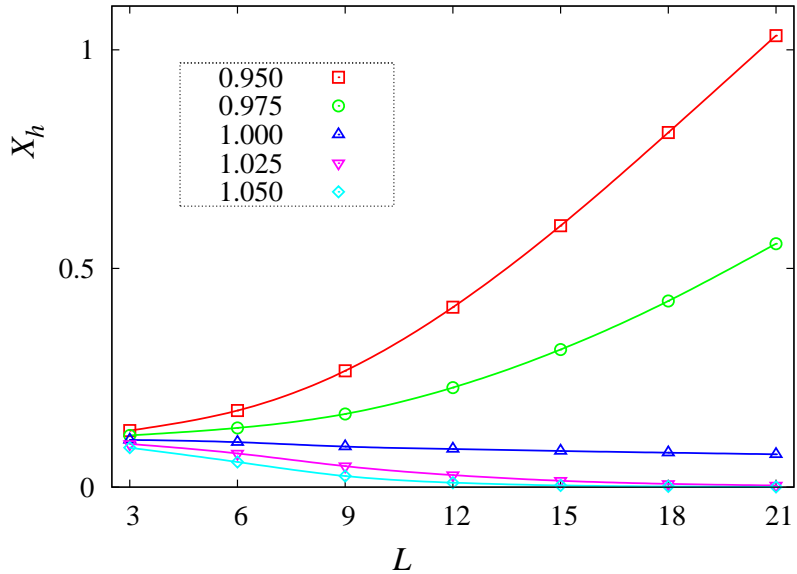}
\caption{(Color online)
Scaled gaps $X_h$ a function of system size $L$, for various pairs of
couplings ($K^{\rm I}_1,K^{\rm I}_2$) proportional to the self-dual
pair with $K^{\rm I}_1=1$ (and  $K^{\rm I}_1=0.13617\cdots$). From top
to bottom, the data apply
to 0.95, 0.975, 1, 1.025, and 1.05 times the self-dual couplings.
These data suggest the presence of a phase transition at or near the
self-dual point.}
\label{scgq2}
\end{figure}

\begin{table}
\caption{Results of transfer-matrix calculations of the scaled
magnetic (left hand side) and energy-like (right hand side) gaps at
the self-dual points of the generalized Baxter-Wu model for several 
values of $K^{\rm I}_1$. The second column indicates the type $s$
of the scaled gap, $h$ for $X_h$ and $t$ for $X_t$. The third column
shows the scaled gap for the largest available system size $L=27$.
Its finite-size dependence is indicated in the fourth column as the
difference between the scaled gaps for the two largest finite sizes.
The effective exponent $p$ describing the finite-size dependence of
the scaled gap is listed in the rightmost column, based on the
scaled gaps for $L=21$, 24, and 27. Positive values of $p$ mean
that the system is renormalizing away from a fixed point.
The values of $X_h$ and $X_t$ in the first line in this table are close
to the exact values of the scaling dimensions; the other entries for
$X_h$ and $X_t$ have no physical meaning except describing the
crossover to another fixed point, possibly with $X_h=X_t=0$.} 
\label{tab_1}
\begin{center}
\begin{tabular}{|l||c|l|r||c|l|r|}
\hline
$K^{\rm I}_1$& $X_h(27)$ & $X_h(27)-X_h(24)$ & $p$  
             & $X_t(27)$ & $X_t(27)-X_t(24)$ & $p$     \\
\hline
0.4407       & 0.124980  & $~0.0000058$      & $-2.2$~ 
             & 0.500626  & $-0.00017  $      & $-2.0$~ \\
0.5          & 0.124412  & $~0.0000051$      & $-2.3$~ 
             & 0.493960  & $-0.00019  $      & $-2.0$~ \\
0.6          & 0.121082  & $-0.000022 $      &   0.47  
             & 0.457856  & $-0.00047  $      & $-1.0$~ \\
0.7          & 0.114513  & $-0.00018  $      &   0.26  
             & 0.398625  & $-0.0018   $      & $-0.23$ \\
0.8          & 0.103767  & $-0.0013   $      &   0.45  
             & 0.324382  & $-0.0044   $      & $ 0.10$ \\
0.9          & 0.088070  & $-0.0017   $      &   0.62 
             & 0.244713  & $-0.0080   $      & $ 0.24$ \\
1.0          & 0.068388  & $-0.0033   $      &   0.83 
             & 0.170464  & $-0.011    $      & $ 0.21$ \\
1.1          & 0.048182  & $-0.0046   $      &   0.42  
             & 0.110422  & $-0.013    $      & $ 0.00$ \\
1.2          & 0.031335  & $-0.0051   $      &   0.00   
             & 0.067915  & $-0.012    $      & $-0.37$ \\
\hline
\end{tabular}
\end{center}
\end{table}
These data show that the scaled gaps for the larger system sizes 
tend to move away from the exact values for the Baxter-Wu
model when  $K^{\rm I}_1$ increases.
Moreover, the finite-size dependence, as indicated by the
difference of the scaled gaps for the two largest system sizes,
increases with $K^{\rm I}_1$, except for the entry for $X_t$ at
largest value $K^{\rm I}_1=1.2$. Another significant phenomenon is
that the exponent $p$ obtained by the three-point fit for
the largest available system is {\em positive} for a range of
$K^{\rm I}_1$, i.e., there are no longer signs of convergence 
with $L$. Only in the case of $K^{\rm I}_1=1.2$ the exponent
becomes negative at the largest available system size, which is
a sign that the renormalized system is approaching an attractive
fixed point.

The presence of a phase transition can be deduced from the  scaling
behavior of the scaled gaps as a function of temperature. The scaled
magnetic gaps were calculated at couplings equal to 0.95, 0.975, 1, 
1.025, and 1.05 times the self-dual pair ($K^{\rm I}_1,K^{\rm I}_2$) 
with $K^{\rm I}_1=1.0$. These data are shown in Fig.~\ref{scgq2}.
For the smallest coupling and largest values of $L$  the behavior 
tends to become linear as a function of $L$, which corresponds with
a correlation length $\xi_m$ that becomes constant, as expected in
a disordered phase. For the largest couplings, the scaled gap tends
rapidly to zero, which corresponds with a long-range ordered phase.
This crossover with increasing $L$, which is to the high temperature
phase or to the ordered phase for ($K^{\rm I}_1,K^{\rm I}_2$)
smaller or larger than the self-dual pair respectively, confirms 
the presence of a phase transition at the self-dual coupling.

\subsection{Monte Carlo results }
\label{mcrq2}
The evidence that the symmetric Baxter-Wu model ($K_1=K_2$, $q=2$) 
undergoes a second-order phase transition is very solid from the
exact solution, an exact mapping to the O(2) loop model on the 
honeycomb lattice \cite{DSS}, and the existing numerical data. 

Using the aforementioned Swendsen-Wang-type cluster algorithm (version 2),
we simulated the $q=2$ generalized Baxter-Wu model at the self-dual line
with $K_1^{\rm I}=0.6$ and $0.8$. The linear system size $L$ was taken as 
multiples of $6$ in the range $6 \leq L \leq 192$; periodic boundary
conditions were imposed. Several quantities were sampled, including the 
number of satisfied up (down) triangles per site $-E_{\rm u}~(-E_{\rm d})$,
the energy density $E$, the specific heat
$C= L^2( \langle E^2 \rangle - \langle E \rangle^2)$, and 
the squared magnetization, defined in analogy with the $n_{\rm P}$-state 
Potts model as 
\begin{equation}
m_{\rm P}^2 = \frac{1}{n_{\rm P}-1}
\sum_{i=1}^{n_{\rm P}-1} \sum_{j=i+1}^{n_{\rm P}} (\rho_i - \rho_j)^2 \; ,
\end{equation}
where we have divided the satisfied triangles into $n_{\rm P}=q^2$ groups
according to the associated ground states, and $\rho_i$, with
$i=1,n_{\rm P}$, is the density of triangles in the $i$th ground state.

We fitted the $C$ data by
\begin{equation}
C (L) = a + b L^{2- 2X_t} 
\end{equation}
and the $m_{\rm P}^2$ data by
\begin{equation}
m_{\rm P}^2(L) = L^{- 2X_h} (a +b L^{-1}) \; ,
\end{equation}
where $a$ and $b$ are unknown constants. The fits yield $X_t =0.43~(2)$
and $X_h=0.1208~(6)$ for $K_1^{\rm I}=0.6$, and  $X_t =0.30~(3)$ and
$X_h=0.110~(2)$ for $K_1^{\rm I}=0.8$. The results are compatible
with those in Table~\ref{tab_1}.

\begin{figure}
\includegraphics[scale=0.9]{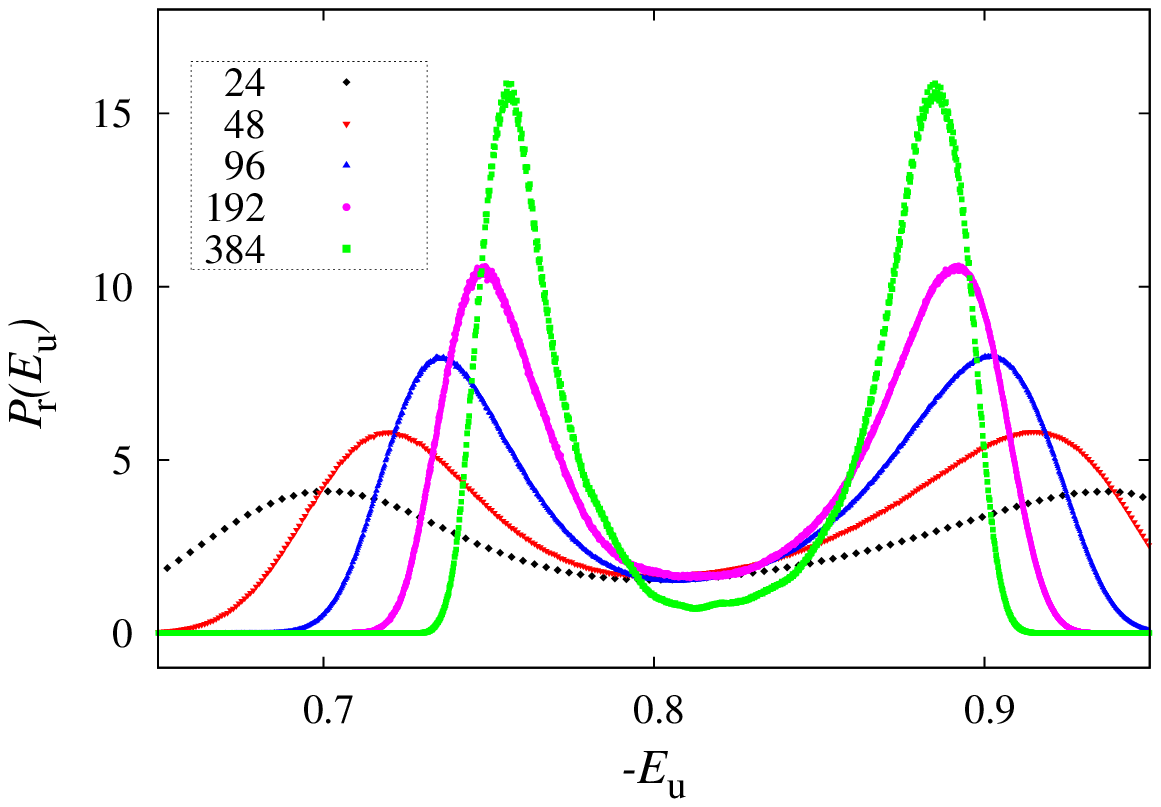}
\includegraphics[scale=0.9]{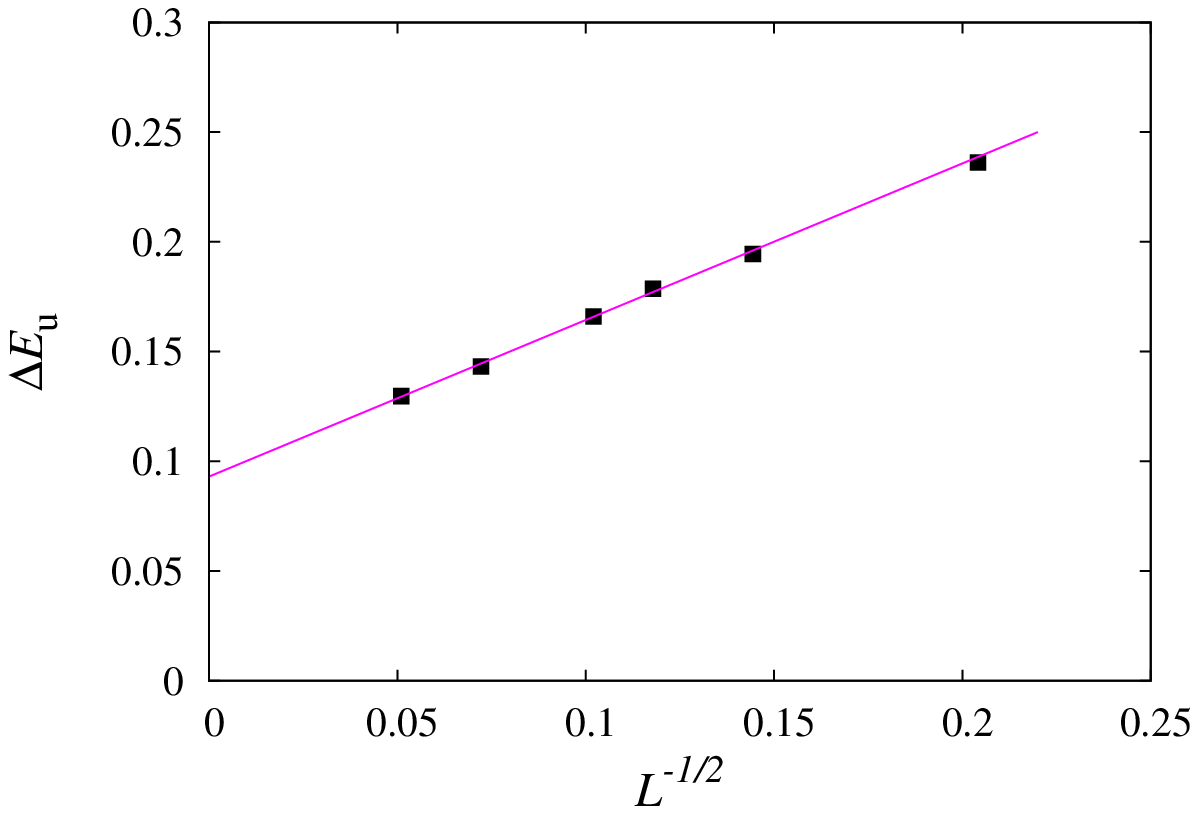}
\caption{(Color online)
Reweighted probability distribution $P_{\rm r}(E_{\rm u})$, and the distance
between the two-peak positions $\Delta E_{\rm u}$ for $K_1^{\rm I}=0.8$.
The height of the peaks increases with system size.}
\label{q2his_Ed}
\end{figure}

The probability distributions $P$ for the sampled quantities are also 
analyzed. The distribution $P(E_{\rm u})$ of the density $-E_{\rm u}$
of the satisfied up-triangles appears to be clearly bimodal, but the
two peaks have unequal heights. 
The reweighted distributions $P_{\rm r}$ were obtained by multiplication
of $P(E_{\rm u})$ with a factor $e^{a+b E_{\rm u}}$, with $a$ and $b$
chosen such that $P_{\rm r}(E_{\rm u})$ is normalized to 1 and that its
two peaks have equal heights. This transformation takes away an overall
gradient in the energy distribution so that the signature of a first
order transition is clearly visible.
Figure~\ref{q2his_Ed} shows  $P_{\rm r}$ as a function of  $E_{\rm u}$,
and the distance $\Delta E_{\rm u}$ between its two maxima.

For first-order transitions, we expect the following behavior of the
reweighted energy distribution:
\begin{enumerate}
\item
The difference between the maximum probability density $\max[P(E_{\rm u})]$
and the local minimum $\min [P(E_{\rm u})]$ between both maxima increases
as $L$ increases \cite{LK};
\item
The distance $\Delta E_{\rm u}$ approaches to a {\em nonzero} value when
 $L \rightarrow \infty$.
\end{enumerate}
The data shown in Fig.~\ref{q2his_Ed} are in agreement with
these conditions. The horizontal scale is chosen as $L^{-1/2}$ because
$\Delta E_{\rm u}$ then behaves approximately linearly in the pertinent
range $24 \leq L \leq 384$. For larger $L$ we expect a faster type of
convergence, which means that the extrapolation in Fig.~\ref{q2his_Ed} 
may slightly underestimate the energy discontinuity for $L \to \infty$.
We also sampled the probability distribution of the 
magnetization-like quantity $m_{\rm P}^2$, and found the same type of
behavior, in agreement with both conditions.
In short, the evidence shown in Fig.~\ref{q2his_Ed} 
for the generalized $q=2$ Baxter-Wu model with $K_1^{\rm I}=0.8$ is just
as expected for a first-order transition.

\begin{figure}
\includegraphics[scale=1.0]{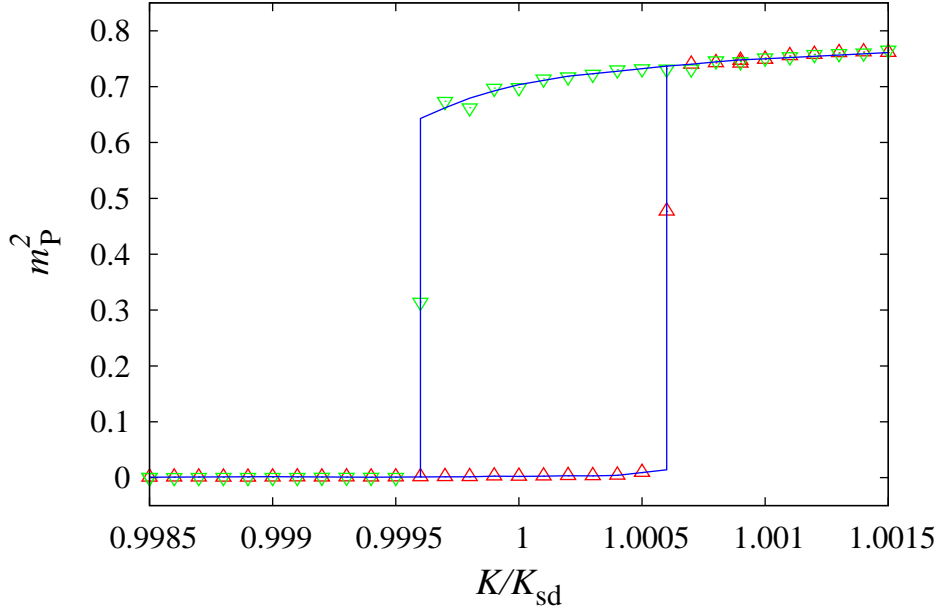}
\caption{(Color online) Hysteresis loop for the squared magnetization
$m_{\rm P}^2$ of an $L=576$ system with $K_1/K_2=5$. The horizontal scale
shows the couplings in units of the self-dual couplings for this ratio.
Each data point represents a simulation of $5\times 10^5$ Metropolis sweeps.
The results for increasing couplings are shown as $\uptr$,
for decreasing couplings as $\dntr$.
The lines are added for visual aid only. }
\label{q2hys}
\end{figure}

In the case of a first-order transition, we also expect metastable
phases in a temperature range about the self-dual point, with  
lifetimes that are much larger than the time scale describing the 
jump from a metastable to a stable branch. We checked for such
hysteresis in the model with $K_1/K_2=5$ by simulations sweeping
slowly over ranges of couplings including the self-dual point.
To find clear hysteresis loops, one has to simulate rather large
systems. Results for $L=576$, with data points representing simulations
of a half million Metropolis sweeps, separated by steps of $10^{-4}$
times the self-dual coupling, are shown  in Fig.~\ref{q2hys}. The
hysteresis loop covers only $10^{-3}$ of the $K/K_{\rm sd}$ scale,
where $K_{\rm sd}$ denotes the self-dual couplings.

\section{Results for $q>2$ }
\label{qgt2}
\subsection{Transfer-matrix calculations}
\label{qgt2TM}
We have constructed transfer-matrix algorithms for the $q=3$ and 4
generalization of the Baxter-Wu model with  $K_1 = K_2$. 
The program is rather similar to that for the Baxter-Wu model, the
main difference is that we have to use ternary or quaternary numbers
to characterize a row of site variables, instead of binary numbers.
As a consequence, a smaller range of system sizes can be handled.
The finite-size data are here restricted to $L \leq 18$ for $q=3$
and $L \leq 12$ for $q=4$.

We computed the largest eigenvalue of the transfer matrix, as well
as the magnetic eigenvalue, characterized by the antisymmetry under
a lattice reflection of the corresponding eigenstate. 
Next, the correlation length and the scaled gap were obtained from
Eqs.~(\ref{calxi}) and (\ref{calsg}). The results for the scaled gap
are shown in Table \ref{tab_2}.

\begin{table}
\caption{
Results of transfer-matrix calculations of the scaled 
magnetic gaps for the $q=3$ and 4 generalized Baxter-Wu models, 
at the respective self-dual points with $K_1=K_2$.
The columns under "$p$" show the exponent obtained
from the three-point fits described in the text.}
\label{tab_2}
\begin{center}
\begin{tabular}{|l||l|l||l|l|}
\hline
\multicolumn{1}{|c||}{}
&\multicolumn{2}{ c||}{ $q=3$ }
&\multicolumn{2}{ c|}{ $q=4$ } \\
\hline
$L$& $X_h(L)$ &  $p$  & $X_h(L)$ &  $p$    \\
\hline
3  & 0.129163 &  ~~~~ & 0.13050 &  ~~~~ \\
6  & 0.117738 &  1.19 & 0.10381 &  1.04 \\
9  & 0.105105 &  0.71 & 0.07655 &  0.37 \\
12 & 0.093650 &  0.62 & 0.05460 &  ~~~~ \\
15 & 0.083255 &  0.54 & ~~~~~~~ &  ~~~~ \\
18 & 0.073778 &  ~~~~ & ~~~~~~~ &  ~~~~ \\
\hline
\end{tabular}
\end{center}
\end{table}
The behavior of the scaled gaps does not suggest convergence
with increasing $L$. Three-point fits according to Eq.~(\ref{3pfit})
yield {\em positive} values of the exponent $p$. This does not agree
well with the description of the finite-size data in terms of an
attractive critical fixed point. It rather suggests crossover to
some other, sufficiently remote fixed point. That may well be a
discontinuity fixed point \cite{NN}. Both for $q=3$ and 4, the
behavior of the scaled gaps as a function of $L$ is similar to that
found in Sec.~\ref{2coupTMres} at intermediate values of $K^{\rm I}_1$.

\begin{figure}
\includegraphics[scale=0.9]{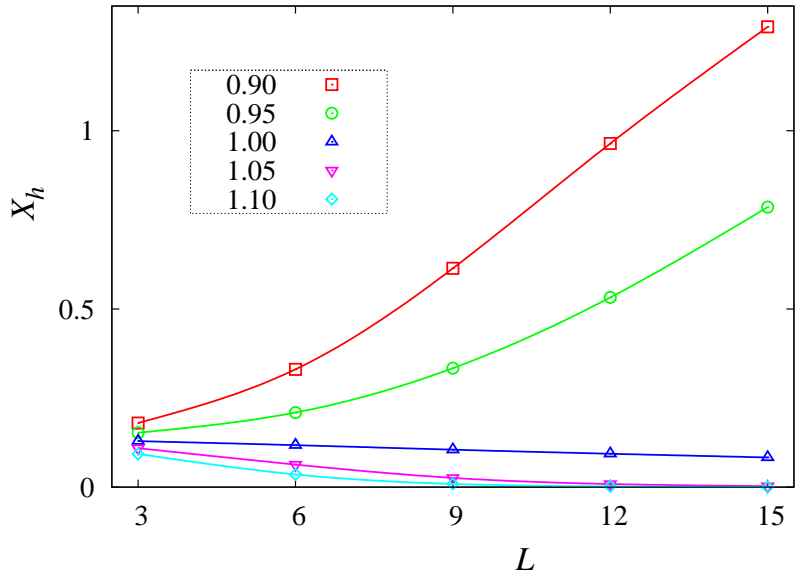}
\caption{(Color online)
Scaled gaps of the $q=3$ generalized Baxter-Wu  model as a function of
system size $L$, for five different couplings
in the vicinity of the self-dual point. From top to bottom, the data 
apply to 0.9, 0.95, 1, 1.05, and 1.1 times the self-dual coupling. These
data suggest that a phase transition takes place near the self-dual point.}
\label{scgq3}
\end{figure}

\begin{figure}
\includegraphics[scale=0.9]{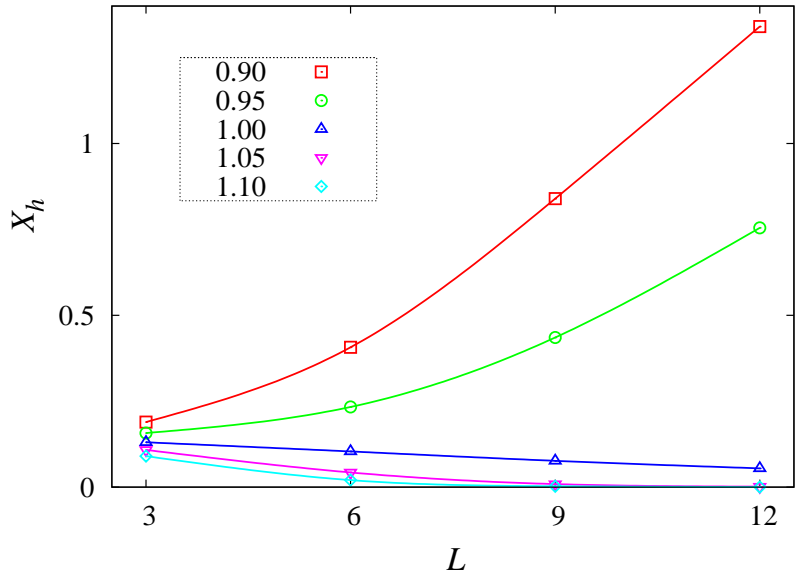}
\caption{(Color online)
Scaled gaps of the $q=4$ generalized Baxter-Wu  model as a function of
system size $L$, for five different couplings
in the vicinity of the self-dual point. From top to bottom, the data 
apply to 0.9, 0.95, 1, 1.05, and 1.1 times the self-dual coupling. These
results are similar to those for the $q=3$ model, but the gap at the
self-dual point decays more rapidly with $L$ for $q=4$.}
\label{scgq4}
\end{figure}

Transfer-matrix calculations at couplings with $K_1=K_2$ in the 
vicinity of the self-dual value show clear signs of transitions.
The scaled magnetic gaps shown in Figs.~\ref{scgq3} and \ref{scgq4}
for $q=3$ and 4 respectively, display the same type of transition
behavior as found  ins Sec.~\ref{2coupTMres} for a $q=2$ model: for
couplings exceeding the self-dual value the scaled gaps tend to zero,
and at the high-temperature side the scaled gaps are increasing with 
system size.

\subsection{Monte Carlo results}
\label{qgt2MC}
Also in this case we employ Monte Carlo simulations to obtain
independent and additional evidence about the character of the
phase transitions. In addition to the evidence already reported
by Alcaraz et al.~\cite{AJ,ACO}, it remains to be investigated
whether hysteresis is present, and whether one can extrapolate the
energy discontinuity to the thermodynamic limit.

We employed the Metropolis method as well as the cluster algorithm 
defined in Sec.~\ref{2coupMC}.
However, in the present case $q > 2$, the efficiency of the cluster
method is not much different from that of the Metropolis algorithm.

\begin{figure}
\includegraphics[scale=0.9]{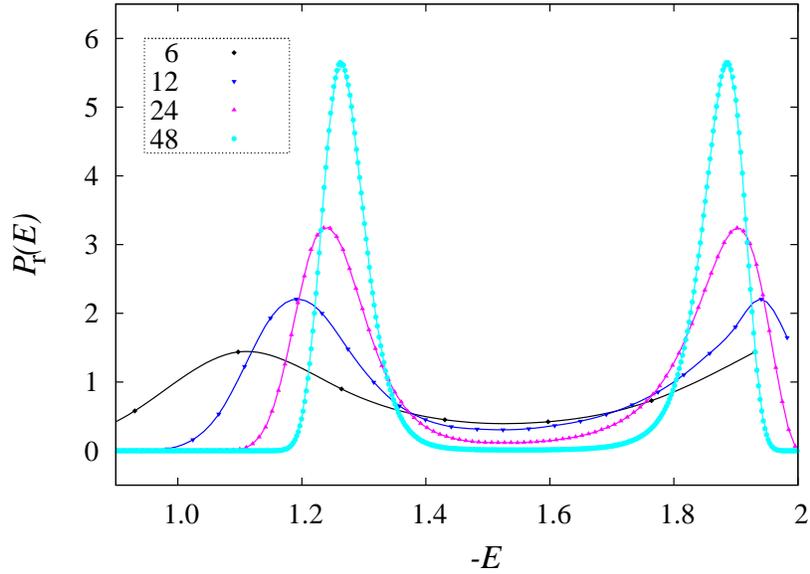}
\caption{(Color online)
Reweighted probability distribution $P_{\rm r}(E)$ for the $q=3$ model.
Data are shown for system sizes $L=6$, 12, 24 and 48. Data points for the
same system size are connected by a curve for the purpose of clarity.
The heights of the peaks increase with system size.}
\label{q3dis}
\end{figure}

\begin{figure}
\includegraphics[scale=0.9]{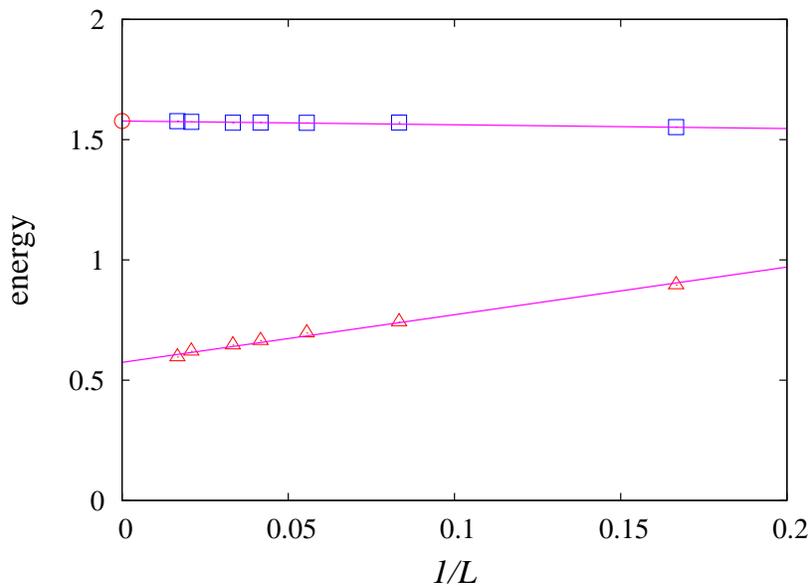}
\caption{(Color online)
Distance along the energy scale between the peaks ($\uptr$), and minus
the mean of the peaks ($\square$) of the energy histogram of the $q=3$
model versus inverse system size. Duality predicts the mean of the
peaks at the position marked by {\hbox{\raise0.7ex\hbox{${}_\bigcirc$}}}.}
\label{ppos3}
\end{figure}

We first simulated the $K_1=K_2$ self-dual point of the $q=3$ model,
and sampled the energy distribution for a number of system sizes that
are multiples of 3. The energy $E$ is defined as minus the density of 
satisfied triangles per site. Again the distribution has two unequal peaks, 
but their separation is wider than in the $q=2$ case.  The reweighting
was done by multiplication of the histogram with $e^{a+b E}$.
The reweighted distribution $P_{\rm r}(E)$ is shown in Fig.~\ref{q3dis}
for several system sizes.
The local minimum between the peaks decreases as a function of $L$.
In the range of finite sizes covered by our simulations, the distance
between the peaks approaches a nonzero constant approximately as $1/L$,
as shown in Fig.~\ref{ppos3}. Such behavior was also found by Lee and
Kosterlitz \cite{LK} for the first-order transition of the $q>4$ Potts
model.  The average of the two peaks, also shown in this figure,
extrapolates within numerical uncertainty to the value $1+1/\sqrt{3}$
predicted by self-duality.

\begin{figure}
\includegraphics[scale=0.9]{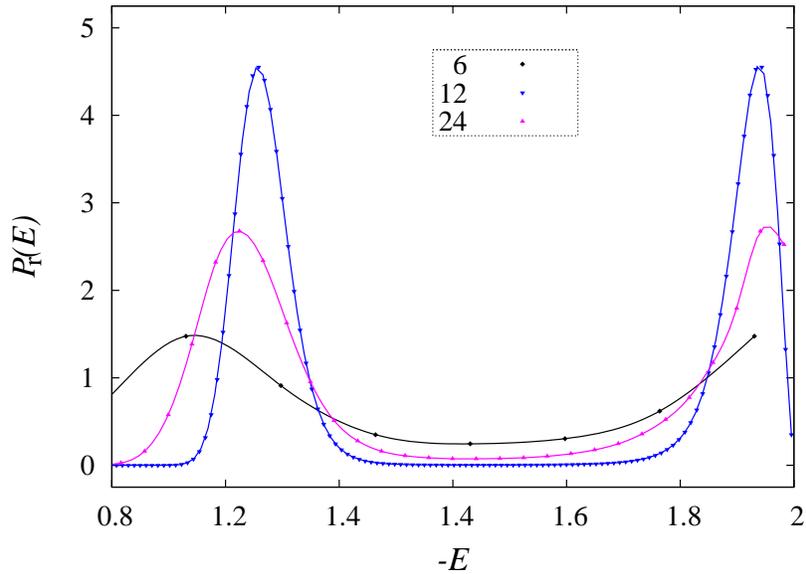}
\caption{(Color online)
Reweighted probability distribution $P_{\rm r}(E)$ for the $q=4$ model.
Data are shown for system sizes $L=6$, 12 and  24. Data points for the
same system size are connected by a curve for the purpose of clarity.
The height of the peaks increases with system size.}
\label{q4dis}
\end{figure}

\begin{figure}
\includegraphics[scale=0.9]{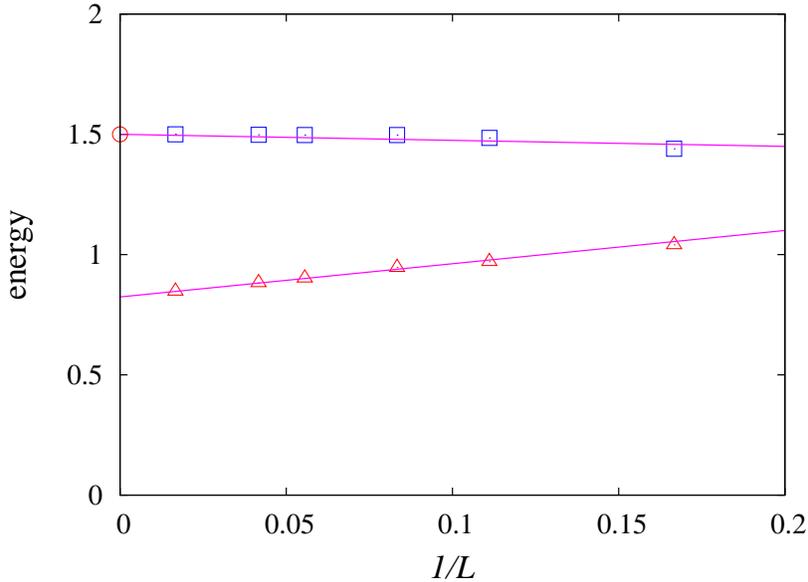}
\caption{(Color online)
Distance along the energy scale between the peaks ($\uptr$), and minus
the mean of the peaks ($\square$) of the energy histogram of the $q=4$ 
model versus inverse system size. Duality predicts the mean of the
peaks at the position marked by {\hbox{\raise0.7ex\hbox{${}_\bigcirc$}}}.}
\label{ppos4}
\end{figure}

Next, we performed similar simulations of the $q=4$ model at the
self-dual point. The reweighted probability distribution $P_{\rm r}(E)$ 
is shown in in Fig.~\ref{q4dis} for several system sizes.
The distances between the maxima of the histogram are shown in
Fig.~\ref{ppos4} as a function of the inverse system size. They
extrapolate to a nonzero constant. The average peak positions, also
shown in Fig.~\ref{ppos3},
agree well with the value $3/2$ predicted by duality.
Also these data agree with the expectations for a first-order transition,
and even more strongly so than in the $q=3$ case, for instance, because
the distances between the peaks of the energy histograms are larger.

\begin{figure}
\includegraphics[scale=1.0]{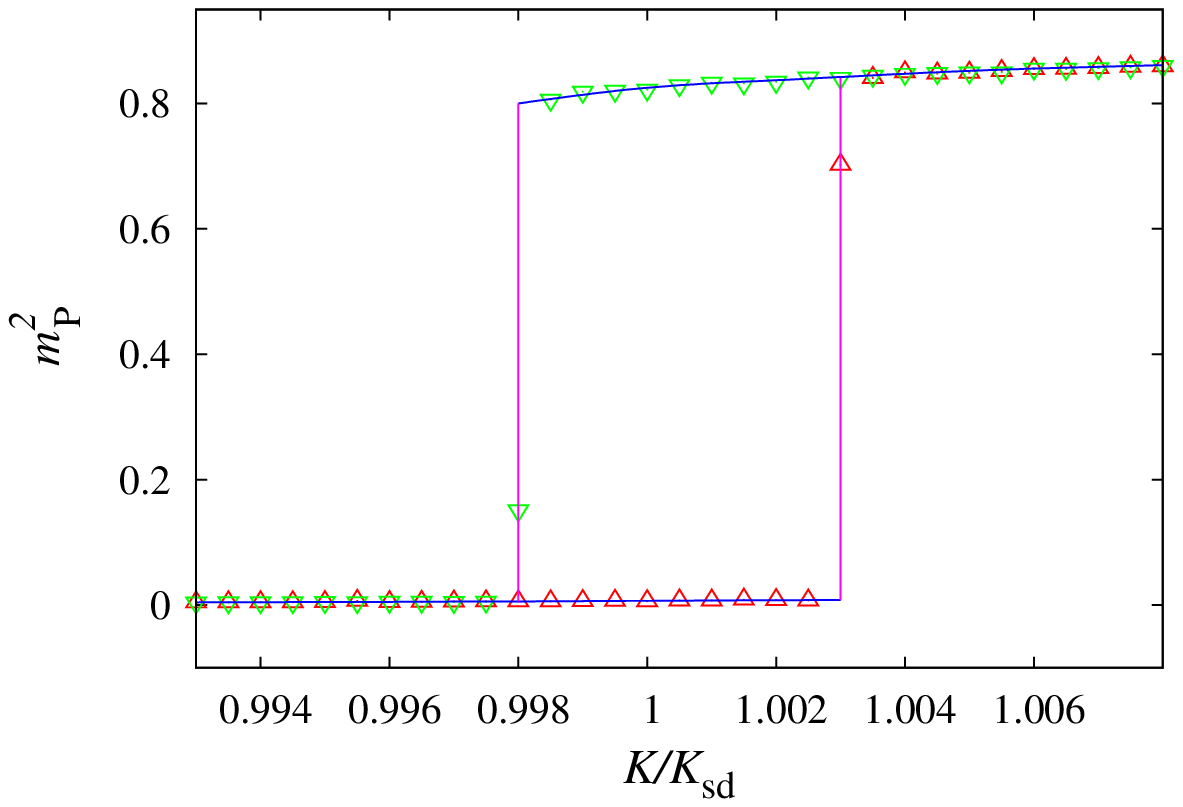}
\caption{(Color online)
Hysteresis loop of the magnetization-like quantity $m^2_{\rm P}$
for the $q=3$ model with size $60^2$. The horizontal scale shows the
coupling in units of the self-dual coupling.  The results for increasing
couplings are shown as $\uptr$, for decreasing couplings as $\dntr$.
The lines are added for visual aid only.  }
\label{q3en}
\end{figure}

\begin{figure}
\includegraphics[scale=1.0]{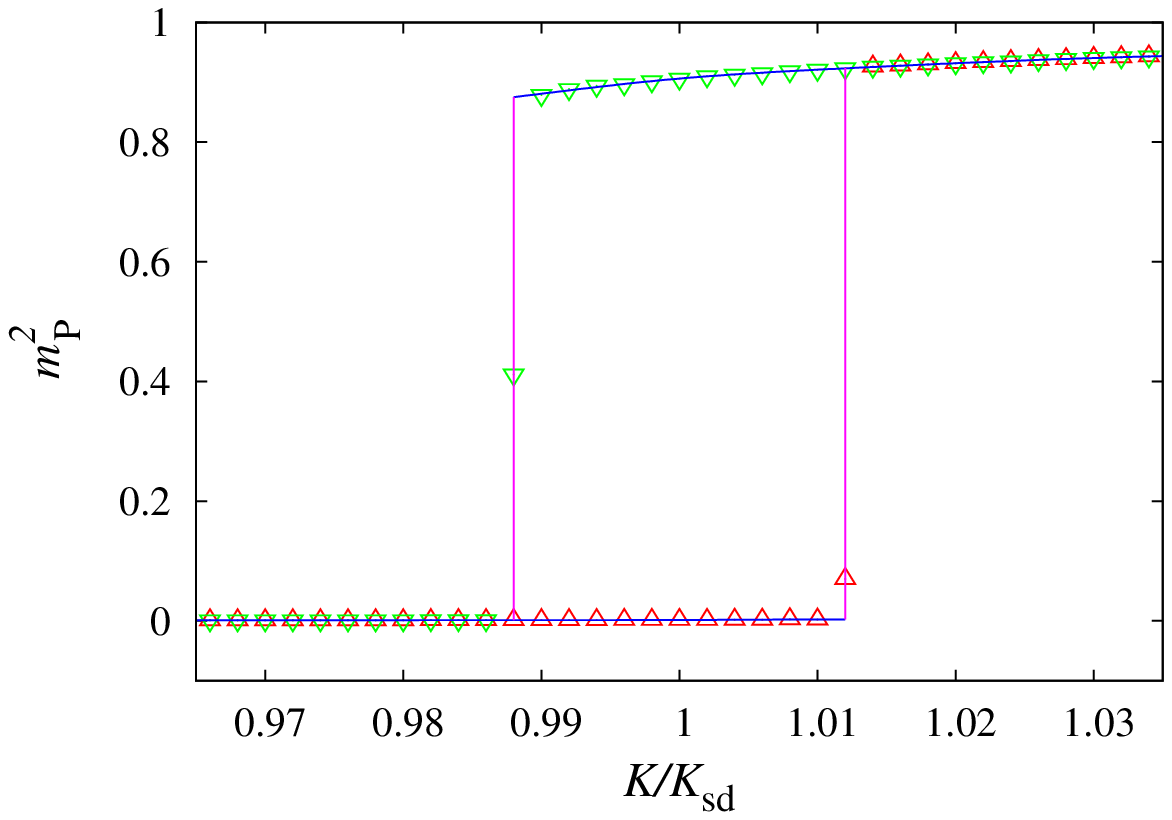}
\caption{Hysteresis loop of the magnetization-like quantity $m^2_{\rm P}$ 
for the $q=4$ model with size $60^2$. The horizontal scale shows the
couplings in units of the self-dual coupling. The results for increasing
couplings are shown as $\uptr$, for decreasing couplings as $\dntr$.
The lines are added for visual aid only.  }
\label{q4en}
\end{figure}

To test for the presence of
hysteresis, we performed Monte Carlo simulations of the $q=3$ and 4 
models, varying the temperature in a region close to the symmetric
self-dual point.  Each data point involved a simulation of $2\times 10^5$
Metropolis sweeps, of which the first $10^4$ were used for equilibration.
The results for the magnetization-type quantity $m^2_{\rm P}$ are shown
in Figs.~\ref{q3en} and Fig.~\ref{q4en}. They
display a small hysteresis loop for $q=3$, covering only a half percent
of the $K$ scale, and stronger hysteresis effects for $q=4$.

\section{Conclusion}
\label{sec_con}
The numerical results presented in Sec.~\ref{2coupTMres} for the
Baxter-Wu model ($q=2$, $K_1 = K_2$) clearly converge to the known
exact values $X_t=1/2$ and $X_h=1/8$.
For $K_1\ne K_2$ deviations from this behavior
are observed, and the dependence of these estimates on the finite
size $L$  is considerable when $K_1$ and $K_2$ are sufficiently
different. At first sight, this situation may seem similar to the
poor convergence observed for some models in the 4-state Potts
universality class, see e.g. Ref.~\cite{BN1982}.

However, there are also significant differences. First we note that,
except for ratios $K_1/K_2$ close to 1, the differences in the
finite-size estimates for $X_t$ and $X_h$ tend to {\em increase} with
increasing system size. Second, the finite-size estimates for  $X_t$
and $X_h$ are {\em smaller} than the exact values for the Baxter-Wu
model, instead of larger as observed for the $q=4$ Potts
model \cite{BN1982,NB83}. 

The interpretation of these observations is suggested by the
renormalization flow diagram for the surface of phase transitions of
the dilute two-dimensional Potts model proposed by Nienhuis
et al.~\cite{NBRS}.
The parameter space of that work involved the chemical potential $v$
of vacant sites and the number of Potts states $q$. The mapping of the
Potts model onto the random-cluster model \cite{KF} enables one to
treat $q$ as a continuous variable. Since vacant sites
in the Potts model are dual to multisite interactions \cite{Yolanda},
the parameter $v$ may as well be interpreted as a scaling field 
depending on the type of interactions.  At $q=4$, the field $v$ becomes
marginal \cite{NBRS} at the critical point.

\begin{figure}
\includegraphics[scale=1.0]{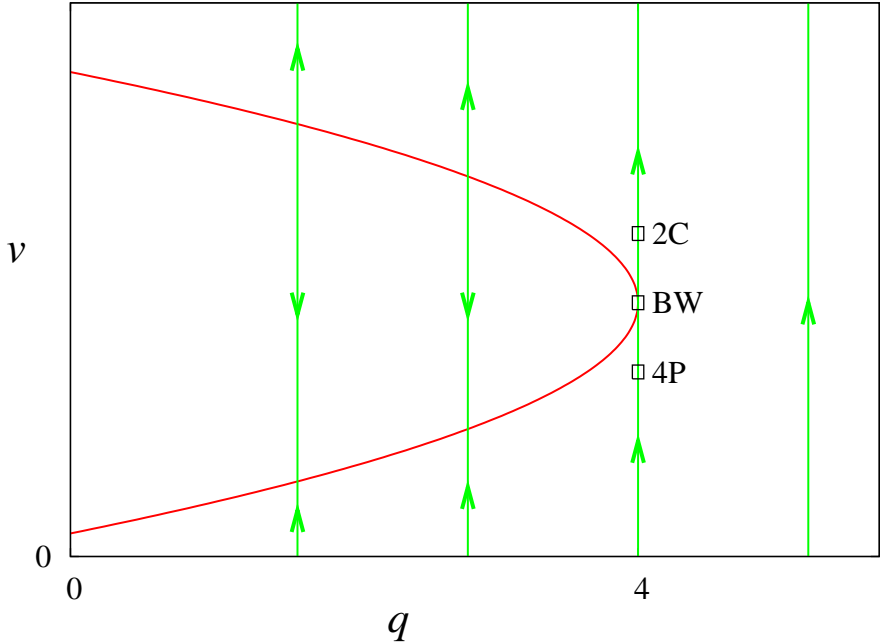}
\caption{(Color online)
Renormalization flow in the plane of phase transitions of the 
dilute random-cluster model, parametrized by the number of Potts
states $q$ and the fugacity $v$ of vacancies, according to
Ref.~\cite{NBRS}. The curve represents a line of fixed points.
Its lower branch is attractive and describes the
critical random-cluster model. The upper branch of fixed points is
repulsive and describes the tricritical random-cluster model. When $v$
exceeds its tricritical value, the renormalization flow leads to a
discontinuity fixed point, corresponding with a first-order transition.
The position of the Baxter-Wu model (BW), of the $q=4$ Potts model
(4P) and the presently investigated self-dual $q=2$ models with
$K_1\ne K_2$ (2C) are sketched.
}
\label{prg}
\end{figure}

We reproduce this flow diagram \cite{NBRS}, adapted to our purposes,
in Fig.~\ref{prg}.  The $q=4$ Potts model is located at a value of $v$
smaller than that at the $q=4$ fixed point, and is still attracted by
it, although marginally. This explains the slow finite-size convergence,
and the logarithmic factors of the  $q=4$ Potts model. The Baxter-Wu
model is located at the $q=4$ fixed point.

The introduction of a difference between $K^{\rm I}_1$ and $K^{\rm I}_2$,
such that the condition of self-duality is still satisfied, allows 
for the possibility that the location of the model in Fig.~\ref{prg}
changes. The coordinate $q$ will remain unchanged, but {\em a priori}
there does not seem to be a way to tell whether the model will move up
or down in the diagram, or perhaps will keep its location.
But, since the finite-size estimates of $X_h$ and $X_t$ for the
$K^{\rm I}_1\ne K^{\rm I}_2$ models and those for the $q=4$ Potts
model lie on opposite sides with respect to the Baxter-Wu model, we 
may locate  the $K^{\rm I}_1\ne K^{\rm I}_2$ models at a value of $v$
exceeding that of the Baxter-Wu model, as indicated by ``2C'' in
Fig.~\ref{prg}.  Therefore they flow to the discontinuity fixed point
\cite{NN} located at large $v$, so that the phase transition is
discontinuous.  In view of the symmetry between $K_1$ and $K_2$,
the marginally relevant field $v$ can, in lowest order, not depend
linearly on $K_1-K_2$ near the 4-state Potts fixed point, and one
expects a contribution as $(K^{\rm I}_1-K^{\rm I}_2)^2$.
This is consistent with the very weak dependence of the finite-size
data in Table \ref{tab_1} on small differences $K^{\rm I}_1-K^{\rm I}_2$.

Thus we conclude that the generalized Baxter-Wu model with different 
couplings described by the Hamiltonian (\ref{BW2c}) undergoes a phase
transition at the self-dual line for $q\geq 2$, and that the
phase transition is first order for $K_1 \ne K_2$, although
extremely weakly so when the difference $K_1-K_2$ is small. Even
for a rather large difference $K_1/K_2=5$, we find (see Fig.~\ref{q2hys})
a very narrow hysteresis loop.

Furthermore, for $q=3$ and 4 the transition is also discontinuous.
This result disproves the possibility mentioned in Sec.~\ref{intro}
that the $q>2$ self-dual generalized Baxter-Wu models renormalize to a
Coulomb gas in which the fugacity of the electric charges vanishes,
in which case algebraic critical behavior would occur.
Apparently the fugacity is nonzero, and, since the electric charges
are relevant for $q>2$, the models renormalize
away from the Gaussian line to a discontinuity fixed point.

The first-order character of the $q=4$ model, as expressed, for instance,
by the energy discontinuity, is stronger than that of the $q=3$ model.
We expect the first-order character to grow even stronger with a further
increase of $q$ and/or the introduction of an asymmetry $K_1 \ne K_2$.

{\em Acknowledgement}:
This research is supported by the NSFC under Grant No.~10675021, and by
the HSCC (High Performance Scientific Computing Center) of the Beijing
Normal University, and, in part, by the Science Foundation of the Chinese
Academy of Sciences.
HB thanks the Beijing Normal University and the University of Science and
Technology of China in Hefei for hospitality extended to him. W. G.
acknowledges hospitality extended to him by the Lorentz Institute.


\begin{thebibliography}{50}
\bibitem{NBRS}
B. Nienhuis, A. N. Berker, E. K. Riedel and M. Schick,
 Phys. Rev. Lett. {\bf 43}, 737 (1979).
\bibitem{NS}
M. Nauenberg and D. J. Scalapino, Phys. Rev. Lett. {\bf 44}, 837 (1980).
\bibitem{BW} R. J. Baxter and F. Y. Wu, Phys. Rev. Lett.
{\bf 31}, 1294 (1973); Aust. J. Phys. {\bf 27}, 357 (1974).
\bibitem{BaN}
G. T. Barkema, M. E. J. Newman, and M. Breeman,
 Phys. Rev. B {\bf 50}, 7946 (1994).
\bibitem{RBP}
R. B. Potts, Proc. Cambridge Philos. Soc. {\bf 48}, 106 (1952).
\bibitem{BN}
B. Nienhuis,  in {\it Phase Transitions and Critical Phenomena},
edited by C. Domb and J. L. Lebowitz. (Academic Press, London, 1987),
Vol. 11, p. 1, and references therein.
\bibitem{AJ} F. C. Alcaraz and L. Jacobs, Nucl. Phys.
B {\bf 210} [FS{\bf 6}], 246 (1982).
\bibitem{ACO}
F. C. Alcaraz, J. L. Cardy and S. O. Ostlund,
J. Phys. A. {\bf 16}, 159 (1983).
\bibitem{KraWa}
H. A. Kramers and G. H. Wannier, Phys. Rev. {\bf 60}, 252 (1941).
\bibitem{GHM} C. Gruber, A. Hintermann and D. Merlini, {\it Group Analysis
of Classical Lattice Systems} (Springer, Berlin 1977).
\bibitem{Turban} L. Turban, J. Phys. C {\bf 15},
L227 (1982).
\bibitem{ZY}
G-M. Zhang and C-Z. Yang, J. Phys. A {\bf 26}, 4907 (1993).
\bibitem{sedua} H. W. J. Bl\"ote, J. R. Heringa and A. Hoogland,
Phys. Rev. Lett. {\bf 63}, 1546 (1989).
\bibitem{MPN}
M. P. Nightingale, Proc. K. Ned. Akad. Wet., Ser. B (Palaeontol.,
Geol., Phys., Chem.) {\bf 82}, 235 (1979).
\bibitem{BN1982} H. W. J. Bl\"{o}te and  M. P. Nightingale,
Physica A (Amsterdam) {\bf 112}, 405 (1982).
\bibitem{QWB}
X.-F. Qian, M. Wegewijs and H. W. J. Bl\"ote,
 Phys. Rev. E {\bf 69}, 036127 (2004).
\bibitem{Cardy-xi}
J. L. Cardy, J. Phys. A {\bf 17}, L385 (1984).
\bibitem{FSS}
For reviews, see e.g. M. P. Nightingale in {\it Finite-Size Scaling and
Numerical Simulation of Statistical Systems}, ed. V. Privman (World
Scientific, Singapore 1990),
and M. N. Barber in {\it Phase Transitions and Critical Phenomena},
eds. C. Domb and J. L. Lebowitz (Academic, New York 1983), Vol. {8}.
\bibitem{NE} M. A. Novotny and H. G. Evertz in {\it  Computer simulation
studies in con\-densed-matter physics} VI, edited by D. P. Landau,
K. K. Mon and H.-B. Sch\"uttler (Springer, Berlin 1993), 188.
\bibitem{Li_89}
X.-J. Li and A. D. Sokal, Phys. Rev. Lett. 63, 827 (1989).
\bibitem{DSS}
Y. Deng, J. Salas and and A. D. Sokal, unpublished (2009).
\bibitem{LK}
J. Lee and J. M. Kosterlitz, Phys. Rev. B {\bf 43}, 3265 (1991).
\bibitem{NN}
B. Nienhuis and M. Nauenberg, Phys. Rev. Lett. {\bf 35}, 477 (1975).
\bibitem{Ferrenberg_88} 
A. M. Ferrenberg and R. H. Swendsen, Phys. Rev. Lett. {\bf  61}, 2635 (1988).
\bibitem{NB83}
M. P. Nightingale and H. W. J. Bl\"{o}te, J. Phys. A {\bf 16}, L657 (1983).
\bibitem{KF}
P. W. Kasteleyn and C. M. Fortuin, J. Phys. Soc. Jpn. {\bf 46}
(Suppl.), 11 (1969);
C. M. Fortuin and P. W. Kasteleyn, Physica (Amsterdam) {\bf 57}, 536 (1972).
\bibitem{Yolanda} Y. M. M. Knops, H. W. J. Bl\"ote and B. Nienhuis,
J. Phys. A {\bf  26}, 495 (1993).
\end{thebibliography}
\end{document}